\documentclass[journal=nalefd,layout=onecolumn,manuscript=article]{achemso}

\usepackage[T1]{fontenc}

\usepackage{setspace}

\usepackage{graphicx}
\graphicspath{{figures/}}
\usepackage{float}
\newfloat{scheme}{htbp}{los}
\floatname{scheme}{Scheme}
\newfloat{chart}{htbp}{loc}
\floatname{chart}{Chart}

\usepackage{amsmath}
\usepackage{bm}
\usepackage[separate-uncertainty = true, per-mode = power]{siunitx}
\usepackage{booktabs}                  
\usepackage[version = 4]{mhchem}       

\usepackage{xcolor} 
\definecolor{darkblue}{rgb}{0.0,0.0,0.55} 
\definecolor{darkred}{HTML}{8B0000}
\usepackage{xr-hyper}
\usepackage[pdftex,plainpages=false,pdfpagelabels,colorlinks=true,citecolor=blue,linkcolor=darkblue,urlcolor=darkred,filecolor=green!50!black,bookmarksopen=true]{hyperref} 
\setkeys{acs}{doi=true}

\usepackage{pdfpages} 



\usepackage{amssymb}
\usepackage{multirow} 

\author{Guilherme B. Kanegae}
 \affiliation{Universidade Estadual de Campinas, Instituto de F\'{i}sica Gleb Wataghin, Departamento de F\'{i}sica Aplicada, Campinas, S\~ao Paulo, 13083-859, Brazil.}
 \email{gkanegae@ifi.unicamp.br}
\author{Alexandre F. Fonseca}
 \affiliation{Universidade Estadual de Campinas, Instituto de F\'{i}sica Gleb Wataghin, Departamento de F\'{i}sica Aplicada, Campinas, S\~ao Paulo, 13083-859, Brazil.}
\email{afonseca@ifi.unicamp.br} 
\title{Beyond Geometry: Hole-Edge Atomic Configuration as a Design Variable in Graphyne Kirigami}

\date{*Email: gkanegae@ifi.unicamp.br; afonseca@ifi.unicamp.br}

\begin{document}

\maketitle

\begin{abstract}

Kirigami-inspired cuts are an established route to stretchability in two-dimensional materials. At the nanoscale, however, cut edges are not purely geometric: they comprise discrete atomic configurations that can alter the mechanical response. In graphene, a given cut geometry fixes the edge structure; graphyne, a porous carbon allotrope with an intrinsic band gap, lifts this constraint, as identical cut geometries can yield distinct hole-edge configurations. Using molecular dynamics simulations, we report the first atomistic study of $\gamma$-graphyne kirigami (GYK), and compared with graphene kirigami. GYK reaches yield strains up to 9.6 times that of pristine $\gamma$-graphyne, with a response governed by both cut geometry and local hole-edge structure. Removing just two carbon atoms at selected edge sites further raises the yield strain by up to 35\%, with little loss of yield stress. These findings identify hole-edge atomic configuration as an additional design parameter, expanding the design space for nanoscale kirigami engineering.

\end{abstract}

\section*{Keywords}
kirigami, graphyne, molecular dynamics, stress-strain, hole-edge configuration.

\bigskip

Graphynes (GYs) are crystalline planar carbon allotropes containing acetylenic chains (\ce{-C#C-})$_n$, $n\geq1$\cite{Ivanovskii2013PSSC}. 
First proposed by Baughman, Eckhardt, and Kertesz in 1987\cite{firstGraphynes}, some GYs can be viewed as graphene in which carbon-carbon bonds are replaced by acetylenic linkages.
GYs are reported to exhibit a direct and tunable electronic bandgap\cite{GYbandgap,GYtunablebandgapStrain,GYtunablebandgapDoping}, high and anisotropic carrier mobility\cite{GYhighcarriermobilitydirectional,GYhighcarriermobility}, and robust mechanical and thermal properties \cite{Fonseca2017DRM,GYreview2019,Douglas2019ACSAMI,Guilherme2022CarbonTrends}. 
These properties make GYs candidates for drug delivery\cite{GYdrugdelivery2025}, hydrogen storage \cite{GYhydrogenStorage,GYhydrogenStorage2}, energy conversion \cite{GYsEnergyConversion,GYsEnergyConversion2,GYsEnergyConversion3}, solid-state cooling \cite{Guilherme2025ACSAMI}, water purification \cite{waterDesalination,waterPermeability,waterPurification}, and catalysis\cite{catalystGY,catalystGY2,catalystGY3}.
Although graphdiyne ($n=2$) and graphtetrayne ($n=4$) were synthesized years ago \cite{synthesisGDY,synthesisG4Y,FonsecaReview2026Carbon}, $\gamma$-GY ($n=1$) has only recently been synthesized by two independent groups \cite{JACSsynthesisGammaGY,CARBONsynthesisGammaGY}.

Despite these attractive properties, $\gamma$-GY, like graphene, is brittle under tensile loading \cite{gyTemperatureSRatefracture2014,gyFractureBehaviors2017}, which limits its applicability in flexible and stretchable devices.
In graphene \cite{Xu2013Nanotech,Liu2025JPCS,Fonseca2026IJMS,Gu2026DiamRelMat} and other two-dimensional (2D) materials\cite{Liu2022MaterTodayNano}, defects such as vacancies, pores, cracks, and grain boundaries have been shown to tune the fracture behavior, in some cases inducing a brittle-to-ductile transition.

A geometric strategy that specifically targets stretchability is kirigami, the Japanese art of paper cutting \cite{Park2019IJPEM,Zhang2020AdvMat,Sun2021Struc,PTAMPES}.
Periodic cut patterns change the dominant deformation mechanism from in-plane stretching to geometric reconfiguration. Out-of-plane deformation modes are activated, local stresses are redistributed, and fracture is delayed, greatly increasing the stretchability relative to pristine membranes.
Kirigami-inspired designs have been used in flexible electronics, metamaterials, morphing structures, and soft robotics \cite{Shyu2015NatMat,kirigamiReview2023,kirigamiEskin,kirigamiFlexiblePiezo,Rubio2023Proc,Lin2025ACSNano,Nagi2025NPJFlexElec,kirigamiMetasurfaces2025}.
Recent examples include platforms for long-term electrophysiological recording\cite{kirigamiNeuralOrganoids2024}, enhanced fog water collection\cite{kirigamiFogNets2023}, and control of the elastocaloric temperature change\cite{Hirai2022AdFuncMat}.

Because its geometric principles are largely scale independent, kirigami has also been applied at the nanoscale. Kirigami patterning enhances stretchability and enables unconventional mechanical functionalities in materials such as graphene \cite{grk2014,grapheneKirigami,HPark2018PRL}, silicon nitride (\ce{SiN}$_x$)\cite{Zhang2020AdvMat}, diamane\cite{zhu2024JNR}, hexagonal boron nitride (\ce{h-BN})\cite{Han2017TSF}, molybdenum disulfide (\ce{MoS2})\cite{Harold2016Nanoscale,Zheng2018CM}, tungsten disulfide (\ce{WS2})\cite{Dey2024RSCAdv}, phosphorene\cite{Gamil2020PLA}, MXenes\cite{Niksan2024NatComm}, and gold\cite{Liu2018SciAdv,Hong2025NatComm}. 
Graphene kirigami (GRK) has become the benchmark platform for understanding geometry-controlled properties, with studies addressing its response to uniaxial tension\cite{grk2014,grapheneKirigami,hua2017SoftMat}, bending\cite{Grosso2015PRL}, shear\cite{grk2025shear}, and out-of-plane loading\cite{Douglas2024CMS}, as well as its thermal \cite{Wei2016Carbon,Bohayra2017Nanoscale}, auxetic \cite{Ho2016PSSB,Cai2016SciRep,Feng2024CMS,Shi2024PRE,Han2025EJMA,Han2025MatTodPhys,Zhang2026CMS}, permeation \cite{Gao2022Carbon}, elastocaloric \cite{Fonseca2023NL,Luis2026NL}, and frictional \cite{Juel2025PNAS} properties.
However, kirigami patterning has not yet been explored in graphyne-based materials.

In contrast to graphene, GYs possess an intrinsic hierarchical structure composed of rigid aromatic rings interconnected by relatively flexible acetylenic linkages\cite{firstGraphynes,Ivanovskii2013PSSC}. 
This arrangement introduces deformation mechanisms that are absent in graphene and leads to mechanical behavior governed by multiple characteristic length scales, as shown by Kanegae and Fonseca for GYs with $1\leq n\leq10$\cite{Guilherme2022CarbonTrends}. 
Kirigami cuts add a second level of structural hierarchy, so the mechanical response is expected to depend on the interplay between the atomic topology and the cut geometry. The atomic boundaries created by the cuts, hereafter referred to as {\it hole edges}, are where these two levels meet, and they can have different atomic configurations for the same cut geometry.
It therefore remains unknown whether the deformation mechanisms established for GRK still hold for GYs and how the atomic configuration of the hole edges affects the mechanical response.

Kirigami engineering of 2D materials has traditionally been viewed as a geometry-driven design problem, in which the mechanical response is controlled by parameters such as cut length, spacing, and density. Here, using molecular dynamics (MD) simulations, we show that this description is incomplete for $\gamma$-GY kirigami (GYK), as the atomic configuration of the hole edges provides an additional and independent degree of freedom. For a given cut geometry, changing the hole-edge configuration modifies the yield strain by up to 29\%, while removing two carbon atoms from selected hole-edge sites produces changes in the yield strain of up to 35\%. Geometric patterning itself increases the yield strain of $\gamma$-GY by up to nearly an order of magnitude relative to the pristine material. Moreover, the effective moduli of GYK are comparable to those of GRK with equivalent geometric parameters, despite the substantially lower intrinsic stiffness of pristine $\gamma$-GY\cite{Lee2008Science,Jiang2009PRB,Guilherme2022CarbonTrends}.
This comparison suggests that the effective stiffness of kirigami nanoribbons is governed primarily by the cut geometry rather than by the parent material, although this mechanism is not directly investigated here. Together, these results identify the local atomic configuration of kirigami hole edges as a new design variable, expanding the design space for tailoring the mechanical properties of graphyne-based nanostructures.

All GYK and GRK structures were constructed and visualized using VMD \cite{vmd}. MD simulations were performed using the open-source LAMMPS package \cite{LAMMPS}. The AIREBO potential \cite{airebo} was employed with the initial REBO cutoff distance set to 2.0 \AA\mbox{}, consistent with previous studies~\cite{Fonseca2019Carbon}. Full details of the simulation setup and computational protocol are provided in the Supporting Information (Section~\ref{SI-sec:methods}).

\begin{figure}[htb!]
  \centering
  \includegraphics[width=\linewidth]{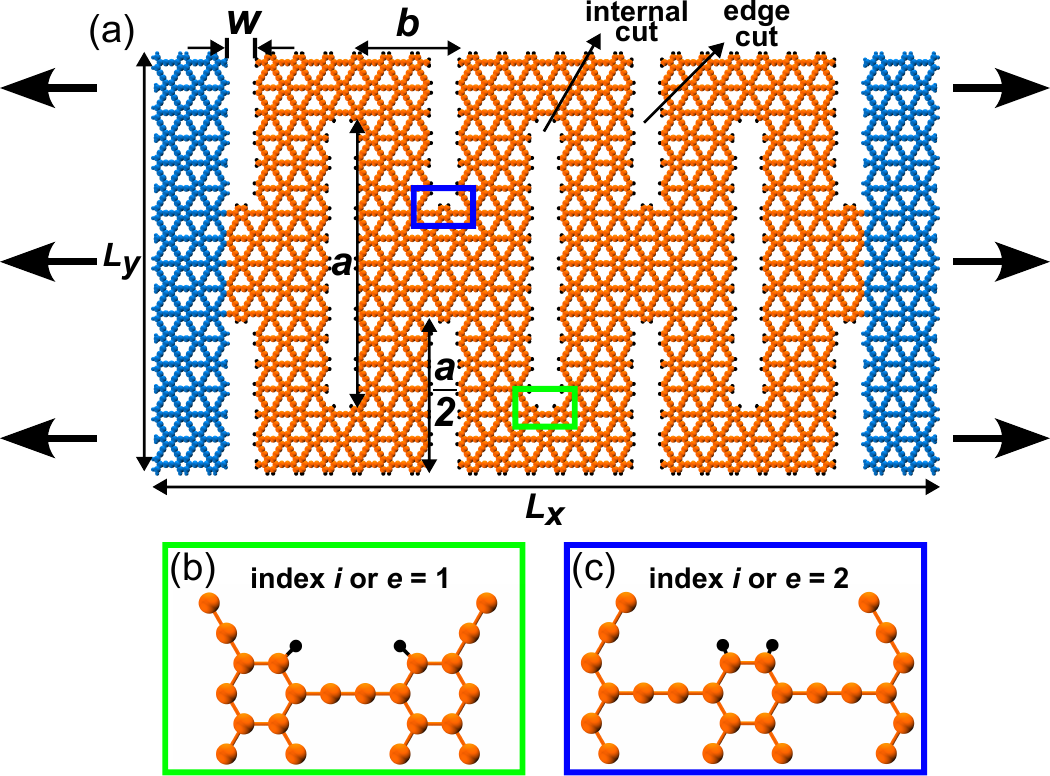}  
  \caption{(a) Kirigami pattern employed in this study with key geometric parameters labeled. C and H atoms are represented in orange and black, respectively. Atoms highlighted in blue were displaced rigidly during simulations. (b, c) Atomic configurations at the hole edges, terminated by (b) an acetylenic linkage or (c) a benzene ring.}
  \label{fig:kirigamistructure}
\end{figure}

The kirigami pattern considered in this study is depicted in Figure~\ref{fig:kirigamistructure}a along with the key geometric parameters. 
The pattern consists of alternating parallel cuts arranged in the interior and at the edges of the nanoribbon. 
The parameter $a$ denotes the length of the interior cuts, while $b$ denotes the distance between adjacent interior and edge cuts. Edge cuts have half the length of the interior cuts ($a/2$), and all cuts share the same width, $w$. 
The total length and width of the nanoribbon are given by $L_x$ and $L_y$, respectively.

When constructing the GYK structures, we found that distinct structures can be obtained from the same set of geometric parameters, a feature arising from the atomic structure $\gamma$-GY provides. 
Specifically, the hole-edge configuration can be chosen such that either an acetylenic linkage or a benzene ring is centered at the top and bottom of each cut, as illustrated in Figures~\ref{fig:kirigamistructure}b and \ref{fig:kirigamistructure}c, respectively. 
To discern between these configurations, the following notation is adopted: GYK$_{ie}$, where 
the index $i$ is associated with the internal cut, and the index $e$ is associated with the edge cut. Each of these indices can assume a value of 1 or 2 (corresponding to structures shown in Figures~\ref{fig:kirigamistructure}b or \ref{fig:kirigamistructure}c, respectively).   
The structure illustrated in Figure~\ref{fig:kirigamistructure}a corresponds to GYK$_{12}$. A total of four configurations were investigated in this study: GYK$_{11}$, GYK$_{22}$, GYK$_{12}$, and GYK$_{21}$. GRK structures were also simulated for direct comparison.

\begin{figure}[htb]
  \centering
  \includegraphics[width=\linewidth]{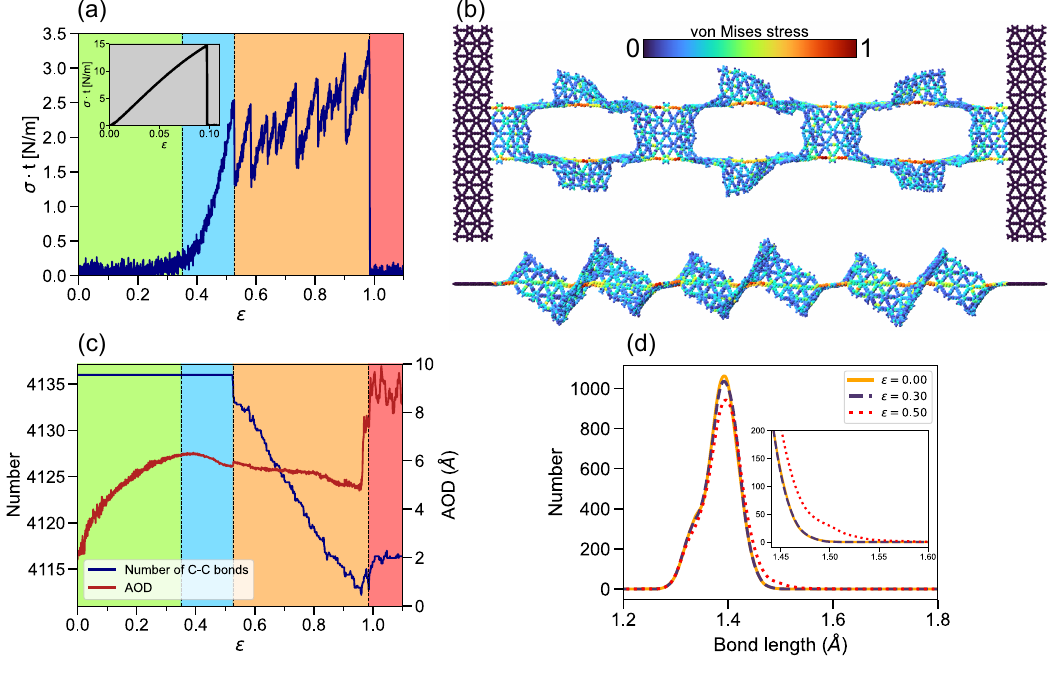}  
  \caption{(a) Stress--strain curve of the representative GYK$_{12}$ structure. The inset shows the stress--strain curve of pristine $\gamma$-GY. (b) Two views of a snapshot of the representative GYK$_{12}$ structure at the end of the second stage of deformation ($\varepsilon$ = 0.53), one from above (top) and the other rotated by $90^{\circ}$ (bottom). Atoms are colored according to the normalized von Mises stress distribution, as indicated by the color scale. (c) Average out-of-plane displacement (AOD) and total number of \ce{C-C} bonds as functions of strain. (d) \ce{C-C} bond length distributions at $\varepsilon=0$, $0.30$, and $0.50$. The inset shows a magnification of the right side of the curves. In (a) and (c), the four deformation stages discussed in the text are delineated by the background colors green, blue, orange, and red.}
  \label{fig:stress-strain-AOD}
\end{figure}

The mechanical response of the GYK$_{12}$ structure under uniaxial tensile loading is presented in Figure~\ref{fig:stress-strain-AOD}a, which shows a representative stress-strain curve. 
We adopted the same background coloring method to distinguish the four deformation regions, as defined in the studies of Qi, Campbell, and Park~\cite{grk2014} and Shi {\it et al}~\cite{grk2025shear}.
The inset in Figure~\ref{fig:stress-strain-AOD}a allows for the comparison of the kirigami response against a pristine $\gamma$-GY nanoribbon of equivalent dimensions. The kirigami pattern dramatically increases the fracture strain (approximately one order of magnitude relative to the pristine material) at the cost of substantially reduced stress throughout the entire deformation range.
Figure~\ref{SI-fig:grk-gr-stress-strain} of the SI displays the stress-strain curve of a representative GRK structure, also showing the results for the pristine graphene structure in the inset.

In the first stage (green region), the cuts flip and rotate to approximately 45$^{\circ}$ relative to the loading direction without appreciable stretching of the \ce{C-C} bonds, resulting in near zero stress. This is confirmed by the bond length distributions shown in Figure~\ref{fig:stress-strain-AOD}d, where the curves at $\varepsilon=0.00$ and $\varepsilon=0.30$ overlap. The increase in average out-of-plane displacement (AOD), observed in Figure~\ref{fig:stress-strain-AOD}c, reflects the rotation of the cuts away from the structure's plane. In the second stage (blue region), bond stretching becomes the dominant mechanism and stress increases until the first bond break occurs. At $\varepsilon=0.50$, the bond length distribution (Figure~\ref{fig:stress-strain-AOD}d) shows a suppressed peak and an increased population at longer bond lengths, while the total bond count in Figure~\ref{fig:stress-strain-AOD}c remains constant, indicating significant bond stretching prior to any fracture.
Figure~\ref{fig:stress-strain-AOD}b shows a representation of the GYK at the end of the second stage of deformation, just before the first bond breaks. As illustrated by the atoms in red, stress concentrates at the borders of internal and edge cuts, while most of the structure remains at low stress values.  

The third stage (orange region) begins with the first bond-breaking event and is characterized by successive partial fractures, as indicated by the steep stress drops in Figure~\ref{fig:stress-strain-AOD}a and the decreasing number of bonds in Figure~\ref{fig:stress-strain-AOD}c. This process culminates in total structural failure in the fourth stage (red region). The sharp increase in AOD during the final deformation stage results from motion of the fractured segments. Each fractured part, no longer held in tension, snaps back toward the fixed boundary to which it remains attached, and this rapid retraction drives large out-of-plane displacements. The slight increase in bond count observed in this region is attributed to bond recombination events that occur when the detached segments fold back toward the rigidly displaced atoms.

To assess the reproducibility of the results, five independent loading simulations were performed on the GYK$_{12}$ and GYK$_{21}$ structures, each initialized with a different random velocity distribution at $T=300$ K.
The corresponding stress-strain curves are shown in Figure~\ref{SI-fig:gyk12-stress-strain-seeds} of the SI. The curves overlap quantitatively up to the first bond break, confirming the reproducibility of this deformation region.
Beyond this point, the stress-strain curves exhibits stochastic behavior, with each simulation reaching total fracture at a different strain value. Given this, the mechanical characterization of the GYKs in this study focuses on the yield strain ($\varepsilon_{\text{yield}}$) and yield stress ($\sigma_{\text{yield}}$), defined as the strain and stress immediately preceding the first bond break, as these quantities are consistent across all simulations.

\begin{figure}[htb!]
  \centering
  \includegraphics[width=\linewidth]{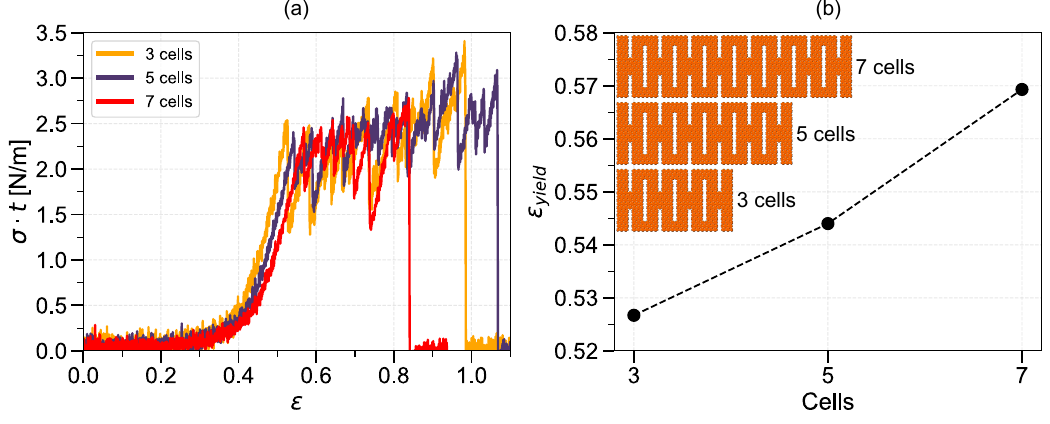}  
  \caption{(a) Stress-strain curves of GYK$_{12}$ structures with varying number of unit cells. (b) Yield strain ($\varepsilon_{\text{yield}}$) as a function of the number of unit cells. The inset shows representative snapshots of the 3-, 5-, and 7-cell structures.}
  \label{fig:lengthdependence}
\end{figure}

The effect of structure length on the mechanical response of the structures was assessed by varying the number of unit cells along the loading direction.
As shown in Figure~\ref{fig:lengthdependence}a, for the GYK$_{12}$ structure, increasing the number of cells produces a systematic shift in the stress-strain curves. 
Figure~\ref{fig:lengthdependence}b shows the corresponding $\varepsilon_{\text{yield}}$ as a function of the number of cells, increasing from 0.527 to 0.569 for structures with 3 and 7 cells, respectively.
This trend differs from GRK, where $\varepsilon_{\text{yield}}$ was reported to be nearly independent of structure length when uniaxial loading was applied \cite{grk2014}. 
Since the dependence with number of cells are not strong for GYKs, for the remainder of this study, structures with 3 cells were used to reduce computational cost.

The mechanical characterization of the GYK structures was performed using two dimensionless parameters, $\alpha$ and $\beta$, defined for GRK systems in previous studies\cite{grk2014,grk2025shear}
\begin{align}
  \alpha &= (a - 0.5L_y)/L_x , \label{eq:a} \\
  \beta  &= (0.5b - w)/L_x , \label{eq:b}
\end{align}
where $a$, $b$, $w$, $L_x$, and $L_y$ are defined in Figure~\ref{fig:kirigamistructure}. Dimensions of all structures are listed in Tables~\ref{SI-tab:geometricparametersALPHAmixed}--\ref{SI-tab:geometricparametersBETA} of the SI.
The parameter $\alpha$ represents the ratio between the overlapping cut and sheet lengths and was varied by modifying the geometric parameter $a$ while keeping all other geometric parameters fixed. $\beta$ represents the ratio between the width of the overlapping cut and the length of the sheet and was varied by modifying $b$.
Note that increasing $b$ also increases the sheet length $L_x$ (Table~\ref{SI-tab:geometricparametersBETA}), so structures in the $\beta$ series also differ in $\alpha$. In contrast, $a$ does not affect $\beta$, so $\beta$ remains constant within each $\alpha$ series. Moreover, because different hole-edge configurations require slightly different lengths, $L_x$, we cannot simulate structures with all of them using the same $\alpha$ or even $\beta$ values.
The full stress-strain curves for GYK and GRK structures with varying $\alpha$ and $\beta$ are shown in Figures~\ref{SI-fig:stress-strain-alphas} and \ref{SI-fig:stress-strain-betas} of the SI, respectively.

\begin{figure}[htb!]
  \centering
  \includegraphics[width=\linewidth]{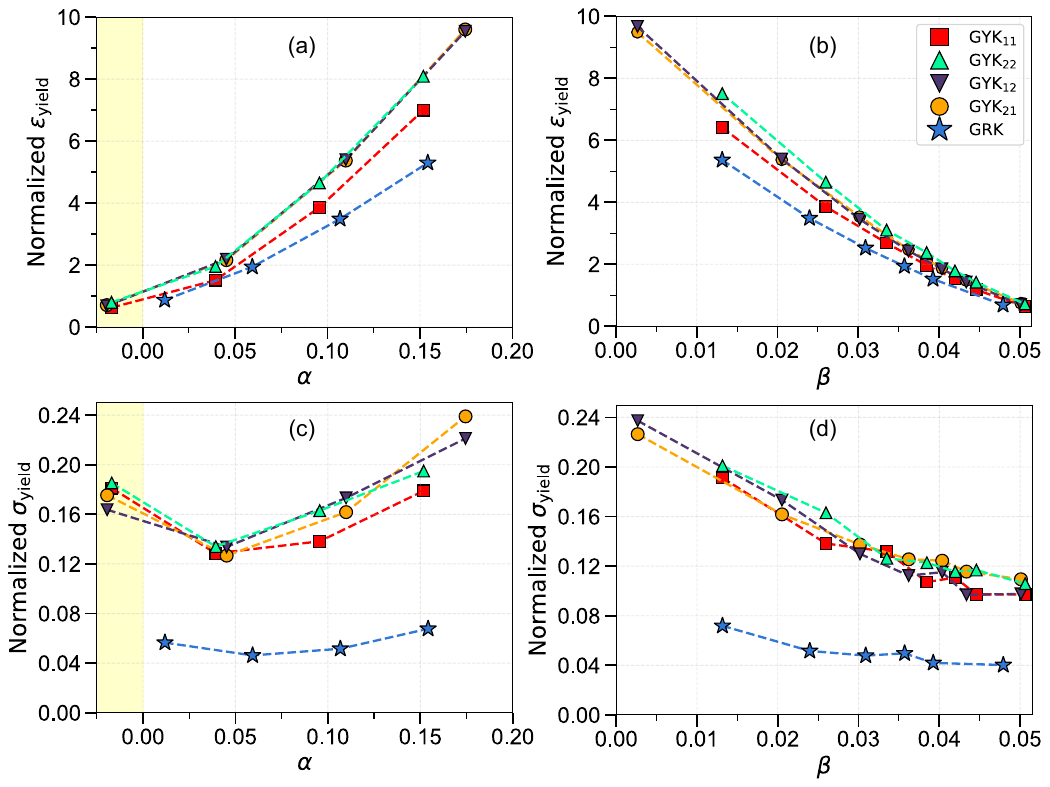}  
  \caption{(a, b) Yield strain ($\varepsilon_{\text{yield}}$) and (c, d) yield stress ($\sigma_{\text{yield}}$) as a function of $\alpha$ (left) and $\beta$ (right). Results are normalized by the corresponding value of its pristine counterparts: $\gamma$-GY for the GYK structures (squares, triangles, and circles) and  graphene for the GRK (stars). The yellow background indicates the region $\alpha<0$, where internal and edge cuts do not overlap and the structure behaves as a cut-free nanoribbon without the deformation mechanism illustrated in Figure~\ref{fig:stress-strain-AOD}b. Dashed lines are guides to the eye. In all panels the error bars are smaller than the symbol size ($\lesssim 0.1$ in (a, b) and $\lesssim 0.002$ in (c, d), see Section~\ref{SI-sec:uncertainty}).}
  \label{fig:yieldstrain-yieldstress}
\end{figure}

Figure~\ref{fig:yieldstrain-yieldstress} displays the dependence of $\varepsilon_{\text{yield}}$ and $\sigma_{\text{yield}}$ with the parameters $\alpha$ and $\beta$. Results are normalized by the corresponding pristine material.
As shown in Figure~\ref{fig:yieldstrain-yieldstress}a, $\varepsilon_{\text{yield}}$ exhibits a nearly linear increase with $\alpha$, reaching a maximum value approximately 9.6 times larger than that of pristine $\gamma$-GY. 
The GYK$_{22}$, GYK$_{12}$, and GYK$_{21}$ configurations display comparable values and trends, as indicated by the overlapping guidelines. 
In contrast, GYK$_{11}$ consistently displays lower $\varepsilon_{\text{yield}}$ values across the entire $\alpha$ range, indicating that this hole-edge configuration, characterized by acetylenic linkages at both hole edges, causes bonds to break at lower strains. 
Because the error bars are smaller than the symbols, this last result for GYK$_{11}$ is a consequence of the different hole-edge configurations among the GYK structures. This suggests the possibility of considering the hole-edge configuration as a new design variable to tune the mechanical properties of GYK structures.
As shown in Figure~\ref{SI-fig:delta-yieldstrain}a of the SI, the difference in $\varepsilon_{\text{yield}}$ between GYK$_{22}$ and GYK$_{11}$ increases with $\alpha$, with GYK$_{22}$ sustaining an additional strain of up to 0.106 before the first bond breaks. In contrast, the difference between GYK$_{12}$ and GYK$_{21}$ remains below 0.006 in magnitude over the entire range of $\alpha$.
The region $\alpha < 0$ (yellow background in Figures~\ref{fig:yieldstrain-yieldstress}a and \ref{fig:yieldstrain-yieldstress}c) corresponds to structures in which the internal and edge cuts do not overlap. In this regime, the deformation mechanism illustrated in Figure~\ref{fig:stress-strain-AOD}b does not occur and the system behaves as cut-free nanoribbons.

A distinct trend is observed for the dependence of $\varepsilon_{\text{yield}}$ with $\beta$, where increasing $\beta$ progressively reduces $\varepsilon_{\text{yield}}$ (Figure~\ref{fig:yieldstrain-yieldstress}b). 
The maximum enhancement remains approximately 9.6 times that of pristine $\gamma$-GY. GYK$_{12}$ and GYK$_{21}$ exhibit nearly identical responses, whereas GYK$_{11}$ consistently shows the lowest $\varepsilon_{\text{yield}}$ among all configurations. 
Conversely, GYK$_{22}$ 
displays a higher $\varepsilon_{\text{yield}}$ than the other structures throughout the investigated $\beta$ range.
Figure~\ref{SI-fig:delta-yieldstrain}b of the SI shows that the difference in $\varepsilon_{\text{yield}}$ between GYK$_{22}$ and GYK$_{11}$ decreases with $\beta$, from 0.107 to 0.008. For GYK$_{12}$ and GYK$_{21}$, the differences are slightly larger than in the $\alpha$ series but remain below 0.018 in magnitude.
Nevertheless, all GYK structures exhibit relatively larger enhancements with respect to the pristine structure than GRK does for the corresponding $\alpha$ and $\beta$ values.

As previously stated, the kirigami geometry significantly increases both yield and fracture strains of the structure, but at the cost of reduced stress throughout the deformation.
The results in Figure~\ref{fig:yieldstrain-yieldstress}c indicate that $\sigma_{\text{yield}}$ generally increases with $\alpha$, although the differences across GYK$_{12}$, GYK$_{21}$, and GYK$_{22}$ configurations are more noticeable than that for $\varepsilon_{\text{yield}}$.
GYK$_{22}$ and GYK$_{12}$ display a more linear dependence, while the remaining structures show less consistent trends. 
Points in the $\alpha < 0$ region are excluded from this trend analysis, as the cut-flipping mechanism does not occur in this regime. 
GYK$_{11}$ exhibits $\sigma_{\text{yield}}$ values comparable to the other structures at low $\alpha$, but falls below them as $\alpha$ increases within the range studied, consistent with the behavior observed for $\varepsilon_{\text{yield}}$.
The lowest recorded $\sigma_{\text{yield}}$ of GYK structures corresponds to a reduction of about 8-fold relative to pristine $\gamma$-GY. In GRK, the largest relative reduction in $\sigma_{\text{yield}}$ is about 2.7 times as large, $\sim 21$-fold.

Figure~\ref{fig:yieldstrain-yieldstress}d exhibits the decreasing trend of $\sigma_{\text{yield}}$  with $\beta$ across all configurations, with an apparent tendency to stabilize for $\beta > 0.05$. 
For $\beta\lesssim0.025$, the trend is well linear for all GYK structures with values of $\sigma_{\text{yield}}$ a little bit larger for GYK$_{22}$ and GYK$_{12}$. For $\beta \gtrsim 0.025$, the values amongst the GYK structures are less distinct. 
Unlike the $\varepsilon_{\text{yield}}$ analysis, no configuration stands out as a clear outlier, with all structures exhibiting comparable values throughout the parameter range.
In this case, the reduction in $\sigma_{\text{yield}}$ ranges from about 4- to 10-fold for GYK, compared with 14- to 24-fold for GRK. 

Following Qi, Campbell, and Park~\cite{grk2014}, we also obtained effective moduli from linear fits to the stress-strain curves. While they refer to these quantities as Young's moduli, we use the term effective modulus, since they characterize the response of the kirigami structure as a whole rather than an intrinsic material property.
$E_1$ corresponds to the first deformation stage (green-shaded regions in Figures~\ref{fig:stress-strain-AOD}a and \ref{fig:stress-strain-AOD}c), while $E_2$ corresponds to the linear portion of the second deformation stage (blue-shaded regions in Figures~\ref{fig:stress-strain-AOD}a and \ref{fig:stress-strain-AOD}c). A detailed account of the fitting procedure, including the determination of the fitting regions, is provided in Section~\ref{SI-sec:determinationeffectivemoduli} of the SI and the strain intervals used to obtain $E_2$ for all structures are illustrated in Figures~\ref{SI-fig:young-regions-grk-alpha}--\ref{SI-fig:young-regions-gyk21}.

\begin{figure}[htb!]
  \centering
  \includegraphics[width=\linewidth]{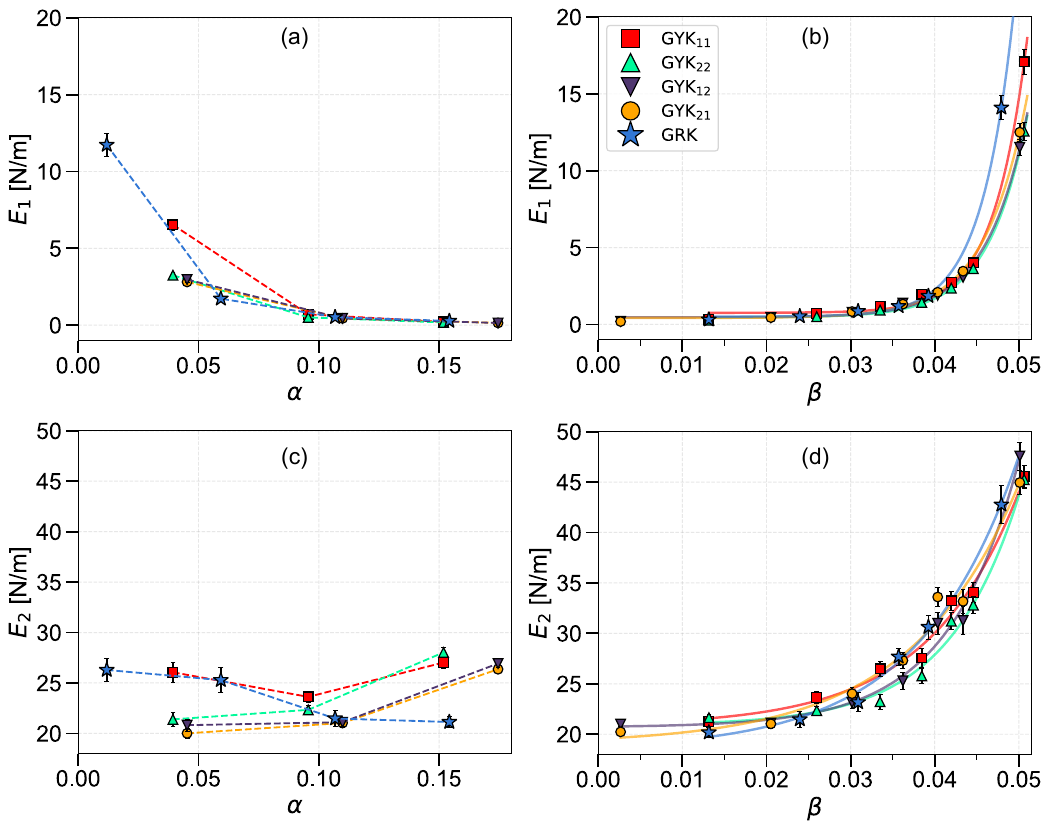}  
  \caption{Effective moduli of the (a, b) first, $E_1$, and (c, d) second, $E_2$, deformation stages as a function of $\alpha$ (left) and $\beta$ (right). $E_1$ was obtained from a linear fit through the origin over the entire first deformation stage and $E_2$ from a linear fit to the region immediately preceding $\varepsilon_{\text{yield}}$ (see Section~\ref{SI-sec:determinationeffectivemoduli}). Error bars correspond to the standard deviation of the fitted slopes. Dashed lines in (a) and (c) are guides to the eye. Solid lines in (b) and (d) are least-squares fits of $E = C_1 \exp(C_2 \cdot \beta) + C_3$ to each structure type.}
  \label{fig:effectivemoduli}
\end{figure}

Figures~\ref{fig:effectivemoduli}a and \ref{fig:effectivemoduli}b show $E_1$ as a function of $\alpha$ and $\beta$, respectively, for all GYK and GRK configurations, while Figures~\ref{fig:effectivemoduli}c and \ref{fig:effectivemoduli}d show the corresponding results for $E_2$.
As shown in Figures~\ref{fig:effectivemoduli}b and \ref{fig:effectivemoduli}d, both $E_1$ and $E_2$ increase with $\beta$, following an exponential dependence well described by $E = C_1 \exp(C_2 \cdot \beta) + C_3$ (solid lines), with $R^2 \geq 0.981$ for all structures. Although an exponential dependence of the effective modulus on cut length has been reported for macroscopic kirigami~\cite{Sun2018APL}, to the best of our knowledge, such behavior has not previously been reported for atomistic kirigami structures. The fitted values of $C_1$, $C_2$, and $C_3$ are listed in Table~\ref{tab:C1C2C3}.

\begin{table}[htb!]
    \caption{The values of $C_1$ [N/m], $C_2$, and $C_3$ [N/m] obtained from fitting the curves of $E_1$ and $E_2$ versus $\beta$ shown in Figures~\ref{fig:effectivemoduli}b and \ref{fig:effectivemoduli}d. The respective coefficients of determination ($R^2$) are also shown.}
    \label{tab:C1C2C3}
	\centering
    \begin{tabular}{lS[table-format=2.0]S[table-format=3.2]S[table-format=1.2]S[table-format=1.3]
    S[table-format=1.2]S[table-format=3.2]S[table-format=2.2]S[table-format=1.3]}
    \toprule
    & \multicolumn{4}{c}{$E_1$} & \multicolumn{4}{c}{$E_2$} \\
    \cmidrule(lr){2-5} \cmidrule(lr){6-9}
    {Structure} & {$C_1 (\times10^{-5})$} & {$C_2$} & {$C_3$} & {$R^2$} & {$C_1$} & {$C_2$} & {$C_3$} & {$R^2$} \\
    \midrule
    GYK$_{11}$ &  5 & 251.06 & 0.75 & 0.997 & 0.23 &  92.30 & 20.80 & 0.988 \\
    GYK$_{22}$ & 22 & 215.45 & 0.46 & 0.999 & 0.06 & 117.83 & 20.87 & 0.990 \\
    GYK$_{12}$ & 36 & 206.36 & 0.46 & 0.998 & 0.07 & 119.04 & 20.68 & 0.988 \\
    GYK$_{21}$ & 53 & 200.18 & 0.42 & 0.999 & 0.54 &  77.43 & 19.00 & 0.981 \\
    GRK        &  6 & 258.53 & 0.50 & 0.999 & 0.43 &  84.44 & 18.44 & 0.995 \\
    \bottomrule
\end{tabular}
\end{table}

The exponential dependence observed for both GYK and GRK points to a strong contribution from the kirigami architecture to the stiffness response. Although $C_1$ and $C_2$ vary among the different hole-edge configurations, potentially indicating that local atomic structure modulates the amplitude and rate of the stiffness increase with $\beta$, these variations do not translate into statistically distinguishable differences in $E_1$ and $E_2$ over the investigated range.
Indeed, the error bars ($\lesssim 1$ N/m, see Tables~\ref{SI-tab:fittingparametersyoungALPHA} and \ref{SI-tab:fittingparametersyoung}) are comparable to or larger than the differences in $E_1$ and $E_2$ between different GYK configurations. Thus, the dependence of the GYK elastic moduli on $\beta$ does no provide sufficient resolution to distinguish the effects of the hole-edge configurations.

The dependence on $\alpha$ is markedly different from that observed for $\beta$. For all GYK and GRK structures, $E_1$ decreases with increasing $\alpha$ (Figure~\ref{fig:effectivemoduli}a). In contrast, $E_2$ exhibits a stronger dependence on the hole-edge configuration (Figure~\ref{fig:effectivemoduli}c): GYK$_{12}$, GYK$_{21}$, and GYK$_{22}$ show an increase in $E_2$ with $\alpha$, whereas GYK$_{11}$ displays a nonmonotonic response, with $E_2$ initially decreasing and then increasing; GRK, in turn, exhibits a monotonic decrease.
Because $\alpha$ is directly related to the cut length $a$, these results indicate that the effect of cut size on the effective modulus depends on the atomic configuration of the hole edges. These trends should nevertheless be regarded as indicative, given that only three to four $\alpha$ values are available for each structure and that $E_2$ is sensitive to the fitting window (see Section~\ref{SI-sec:windowselection} of the SI).
Even with these limitations, the distinct $E_2$ responses among the GYK configurations suggest that the local atomic structure of the hole edges provides an additional degree of freedom for tailoring the mechanical response of 2D kirigami structures.

The effective modulus of the second deformation region of 15.17 N/m reported by Qi, Campbell, and Park~\cite{grk2014} for GRK is lower than our present results. This discrepancy may arise from differences in the MD simulation and fitting protocols employed.

\begin{figure}[htb!]
  \centering
  \includegraphics[width=\linewidth]{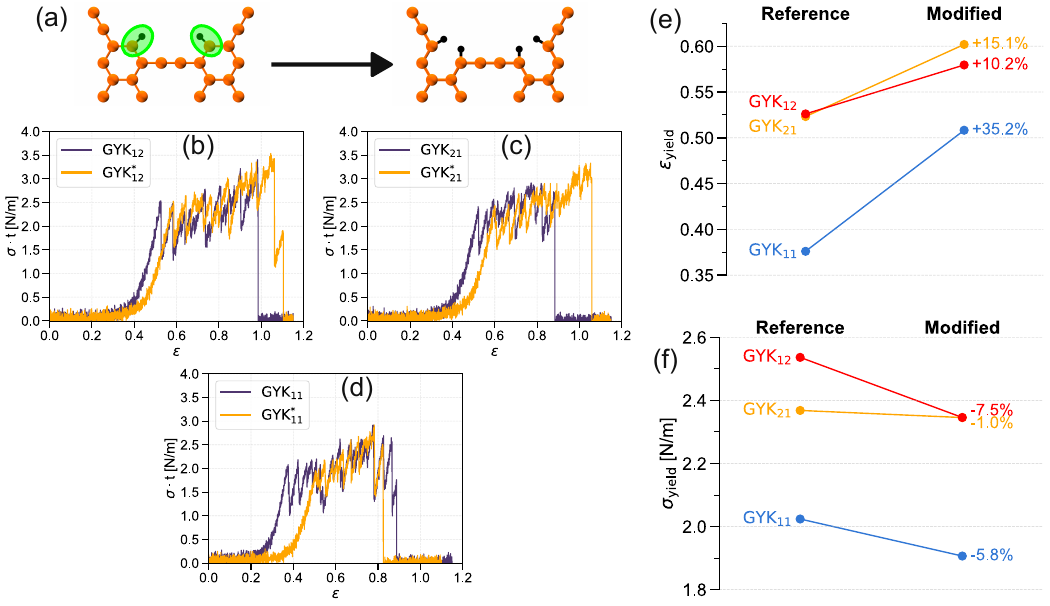}  
  \caption{(a) Atomic configurations at the hole edges before (left) and after (right) the removal of carbon atoms of the benzene ring at hole edge. The modification is applicable to hole edges featuring an acetylenic linkage. Removed atoms are highlighted in green. (b--d) Stress--strain curves of the reference (purple) and modified (orange) (b) GYK$_{12}$, (c) GYK$_{21}$, and (d) GYK$_{11}$ structures. (e, f) Differences in (e) yield strain and (f) yield stress between the modified and reference structures. Labels indicate the corresponding relative differences.}
  \label{fig:atomicmodification}
\end{figure}

A further consequence of locally altering the hole-edge configuration is illustrated in Figure~\ref{fig:atomicmodification}, which shows the effect of removing specific carbon atoms from the hole edges on the stress-strain response. The atomic modification is depicted in Figure~\ref{fig:atomicmodification}a and can be applied only to structures containing at least one hole edge with an acetylenic linkage. GYK$_{22}$ is therefore excluded from this analysis. The modified structures are denoted by an asterisk (*) in the legends of Figures~\ref{fig:atomicmodification}b--d. 
In all cases, removing these atoms shifts the stress-strain curves toward larger strains, extending the near-zero-stress regime and increasing $\varepsilon_{\text{yield}}$.

The relative increases in $\varepsilon_{\text{yield}}$ are shown in Figure~\ref{fig:atomicmodification}e. 
For GYK$_{12}$ and GYK$_{21}$, $\varepsilon_{\text{yield}}$ increases by $10$\% and $15$\%, respectively. The largest enhancement occurs for GYK$_{11}$, in which both the internal and edge cuts contain acetylenic linkages. Applying the atomic modification to both types of cut produces a $35$\% increase in $\varepsilon_{\text{yield}}$.
This extension of the elastic range is accompanied by a reduction in $\sigma_{\text{yield}}$ (Figure~\ref{fig:atomicmodification}f), with decreases of $7.5$\% and $5.8$\%, for GYK$_{12}$ and GYK$_{11}$ respectively, whereas GYK$_{21}$ shows only a $1.0$\% reduction.

The origin of the increase in $\varepsilon_{\text{yield}}$, accompanied by a decrease in $\sigma_{\text{yield}}$, can be understood from the local atomic configuration around the removed atoms, as illustrated in Figure~\ref{fig:atomicmodification}a. 
Removing the carbon atoms highlighted in the left panel creates a hinge-like feature that locally softens the structure by introducing an additional deformation mode, thereby allowing greater extension along the loading direction. Because the acetylenic linkages are intrinsically stiff in tension, bending at these newly formed corners is energetically more favorable than stretching the \ce{C-C} bonds. This local redistribution of deformation accounts for the simultaneous increase in $\varepsilon_{\text{yield}}$, and decrease in $\sigma_{\text{yield}}$.


In this work, we systematically investigated the tensile behavior of $\gamma$-graphyne kirigami (GYK) and compared its mechanical response with that of graphene kirigami (GRK) structures with analogous geometries. 
By independently varying the kirigami parameters $\alpha$ and $\beta$ and considering four distinct hole-edge configurations, we disentangled the effects of global geometry and local atomic structure on the mechanical response of GYK. The normalized $\sigma_{\text{yield}}$ and $\varepsilon_{\text{yield}}$ exhibit the same general geometric trends reported for GRK, increasing with $\alpha$ and decreasing with $\beta$.

Despite these common trends, GYK exhibits a substantially smaller reduction in $\sigma_{\text{yield}}$ upon patterning than its graphene counterpart. Because pristine $\gamma$-GY is intrinsically more compliant than graphene, the relative mechanical penalty associated with introducing kirigami cuts is reduced. As a result, $\gamma$-GY benefits more strongly from kirigami patterning in terms of ductility, exhibiting consistently larger relative enhancements in $\varepsilon_{\text{yield}}$ than graphene.

The effective moduli of both GYK and GRK exhibit the same exponential dependence on $\beta$, indicating that this scaling behavior is primarily governed by the kirigami architecture rather than by the underlying carbon framework. In contrast, the dependence on $\alpha$ varies among the GYK configurations, particularly for $E_2$, revealing a sensitivity to the local atomic structure of the hole edges. These results point to an interplay between global kirigami geometry and local atomic configuration in determining the effective mechanical response.

Beyond geometric effects, our results show that the atomic configuration of the hole edges provides an additional design variable for GYK. Even without changing the overall kirigami geometry, modifying the local structure at cut corners and hole perimeters can substantially alter $\sigma_{\text{yield}}$ and $\varepsilon_{\text{yield}}$.  This sensitivity arises from the mixed sp/sp$^2$ bonding network of $\gamma$-GY, which allows identical cut geometries to generate distinct hole-edge configurations that have no direct counterpart in graphene. The local bonding arrangement therefore provides an additional atomistic degree of freedom for tailoring mechanical performance.

Overall, these findings demonstrate that the mechanical response of GYK cannot be fully described by geometric parameters alone. It emerges from the interplay among kirigami geometry, the intrinsic mechanical properties of the graphyne framework, and the local atomic configuration of the hole edges. By establishing these coupled structure-property relationships, this work expands the atomistic design space of two-dimensional kirigami and provides a framework for rationally tailoring the mechanical response of graphyne-based carbon nanostructures.

\section*{Supporting Information}

The Supporting Information is available free of charge.
\begin{itemize}
  \item \textbf{Supporting Information}: computational details of the molecular dynamics simulations; stress-strain curves from five statistically independent simulations for GYK$_{12}$ and GYK$_{21}$; full stress-strain curves for all GYK and GRK structures and the corresponding structural dimensions; comparison of the yield strain between GYK structures with the same geometric parameters but different hole-edge configurations; procedure used to determine the effective moduli $E_1$ and $E_2$ from linear fits, with the fitting regions and resulting values for each structure; sensitivity of $E_2$ to the fitting window and limitations of $R^2$ as a selection criterion; procedure for estimating the uncertainties of the normalized yield strain and yield stress (PDF).
  \item \textbf{Supporting Files}: AIREBO force field used in the simulations; initial structures (PDB) and ready-to-run LAMMPS data files for the graphyne and graphene kirigami, and pristine nanoribbon structures; equilibration and loading scripts; example simulation outputs; example videos of the equilibration and loading steps for a representative structure (ZIP).
\end{itemize}



\section*{Author Contributions}

G.B.K.\ and A.F.F.\ designed the project. G.B.K.\ generated the structures, ran the MD simulations, and made all figures and graphics. A.F.F.\ supervised the project. G.B.K.\ and A.F.F.\ analyzed the results, wrote, and reviewed the manuscript. 

\section*{Notes}

The authors declare no competing financial interest.

\section*{Acknowledgements}

The authors thank the Brazilian Coordination for the Improvement of Higher Education
Personnel (CAPES) for covering the open access publication costs. This work was partly supported by the Brazilian agencies National Council for Scientific and Technological Development (CNPq) and the São Paulo Research Foundation (FAPESP), Brazil.
G.B.K.\ acknowleges support from FAPESP (grant \#2024/15033-6). A.F.F.\ is a fellow of the Brazilian Agency CNPq (\#302009/2025-6) and acknowledges
grant \#2024/14403-4 from FAPESP. The authors also thank the Coaraci Supercomputer for computer time (FAPESP grant \#2019/17874-0) and the Center for Computing in Engineering and Sciences at Unicamp (FAPESP grant \#2013/08293-7).
The authors used GPT-5.6 Luna (OpenAI) and Claude Sonnet 5.5 (Anthropic) for language editing and improvement of the manuscript. The authors are fully responsible for the scientific content, interpretation, and final wording.

\newpage
\bibliography{references}



\includepdf[pages=-]{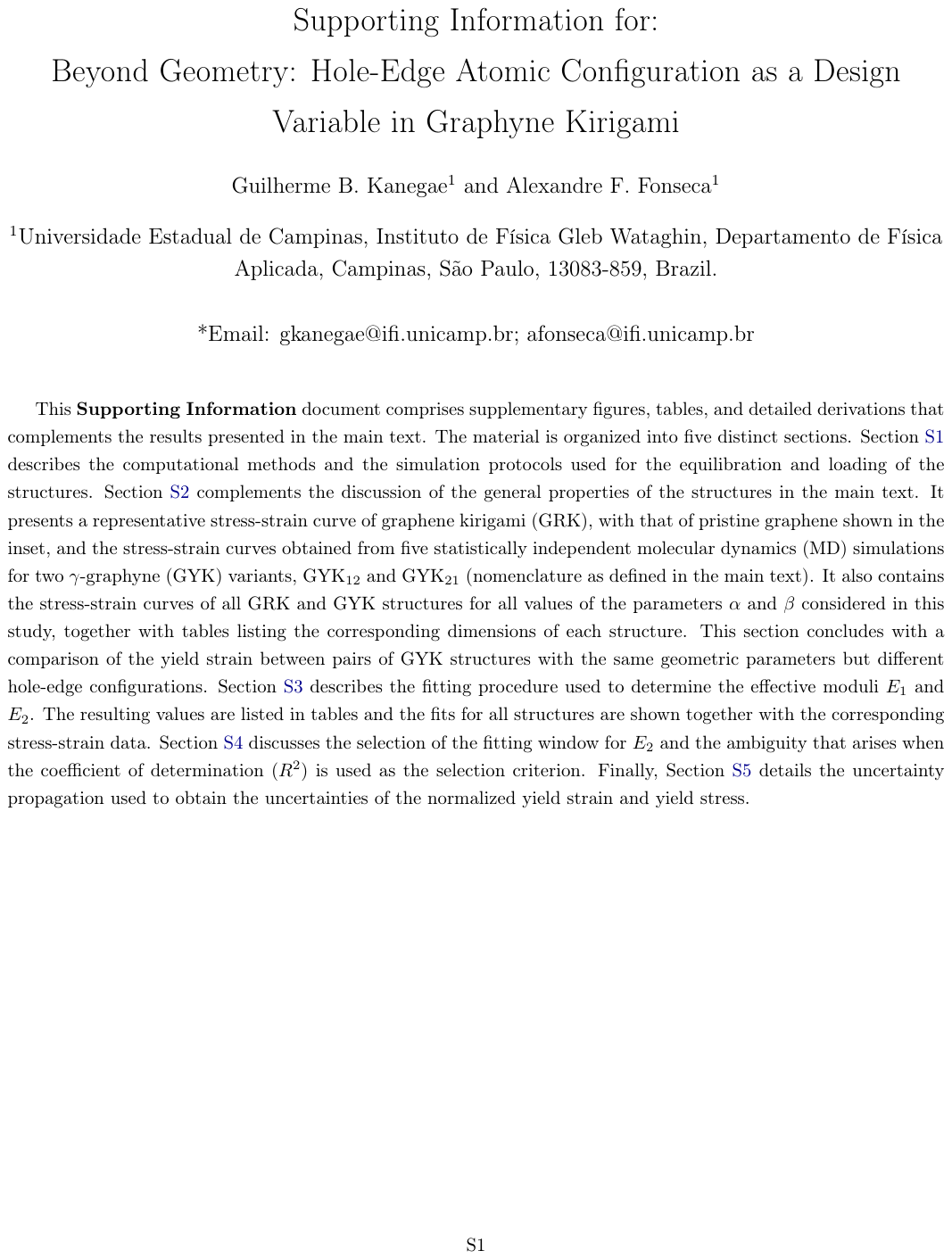}

\end{document}